\documentclass[
 reprint,
 amsmath,amssymb,
 aps,
floatfix,showkeys,
]{revtex4-1}

\usepackage[english]{babel}
\usepackage[utf8]{inputenc}
\usepackage{float}
\usepackage[caption=false]{subfig}
\usepackage{graphicx}% Include figure files
\usepackage{dcolumn}% Align table columns on decimal point
\usepackage{bm}% bold math
\usepackage{color}
\usepackage{wrapfig}
\usepackage{physics}
\usepackage{fancyhdr}
\usepackage{lastpage}

\def\_#1{\textsubscript{#1}}
\def\^#1{\textsuperscript{#1}}
\newcommand{\angstrom}{\mbox{\normalfont\AA}}

\begin{document}

\title{Enhancing two-photon spontaneous emission in rare earths using graphene and graphene nanoribbons} 

\author{Colin Whisler} 

\author{Gregory Holdman}

\author{D. D. Yavuz}

\author{Victor W. Brar}
\affiliation{University of Wisconsin-Madison, Department of Physics, 1150 University Ave., Madison, Wisconsin 53706, USA}
\email{vbrar@wisc.edu}

\begin{abstract}

The enhancement of two-photon spontaneous emission (2PSE) from trivalent and divalent rare earth ions in proximity to graphene and graphene nanoribbons is calculated for achievable experimental conditions using a combination of finite difference time domain simulations and direct computation of transition rates between energy levels in rare earths. For Er\^{3+}, we find that the 2PSE rate is initially 8 orders lower than the single-photon spontaneous emission rate but that, with enhancement, 2PSE can reach 2.5\% of the overall decay. When graphene nanoribbons are used, we also show that the emission of free-space photon pairs from Er\^{3+} at 3 - 3.2 $\mu$m via 2PSE can be increased by $\sim 400$. Our calculations show significantly less relative graphene-enhanced 2PSE than previous works, and we attribute this variation to differences in emitter size and assumed graphene mobility. We also show that the internal energy structure of the ion can have an impact on degree of 2PSE enhancement achievable and find that divalent rare earths are more favorable.

\end{abstract}

\maketitle

\section{Introduction}

Two photon spontaneous emission (2PSE) is a decay process whereby an electronic transition occurs via emission of a photon pair whose combined energy is equal to the difference between the excited and ground states \cite{Goppert-Mayer_1931}. Compared to single-transition decay, 2PSE usually occurs at a much lower rate, as it requires a transition to an intermediate state. In atomic systems these states are typically found at energies far from the initial and final states, and they are only weakly connected through an electric dipole perturbation. This leads to 2PSE decay rates that are commonly $10^8$ lower than single-channel transition rates \cite{Breit_Teller}. Despite this low efficiency, 2PSE has generated interest as a mechanism for the creation of entangled photon pairs that would be useful in quantum information systems \cite{Hayat_Ginzburg_Orenstein_2007}. In particular, the 2PSE process naturally generates photon pairs that are entangled polarization \cite{Perrie_Duncan_Beyer_Kleinpoppen_1985, Radtke_Surzhykov_Fritzsche_2008} and frequency states \cite{Hayat_Ginzburg_Neiman_Rosenblum_Orenstein_2008}. Moreover, the theoretical bounds on 2PSE permit it to, in principle, generate photon pairs at higher rate than state-of-the-art parametric down-conversion schemes, and over a wider range of frequencies.

\begin{figure*}
    \centering
    \includegraphics[scale=0.32]{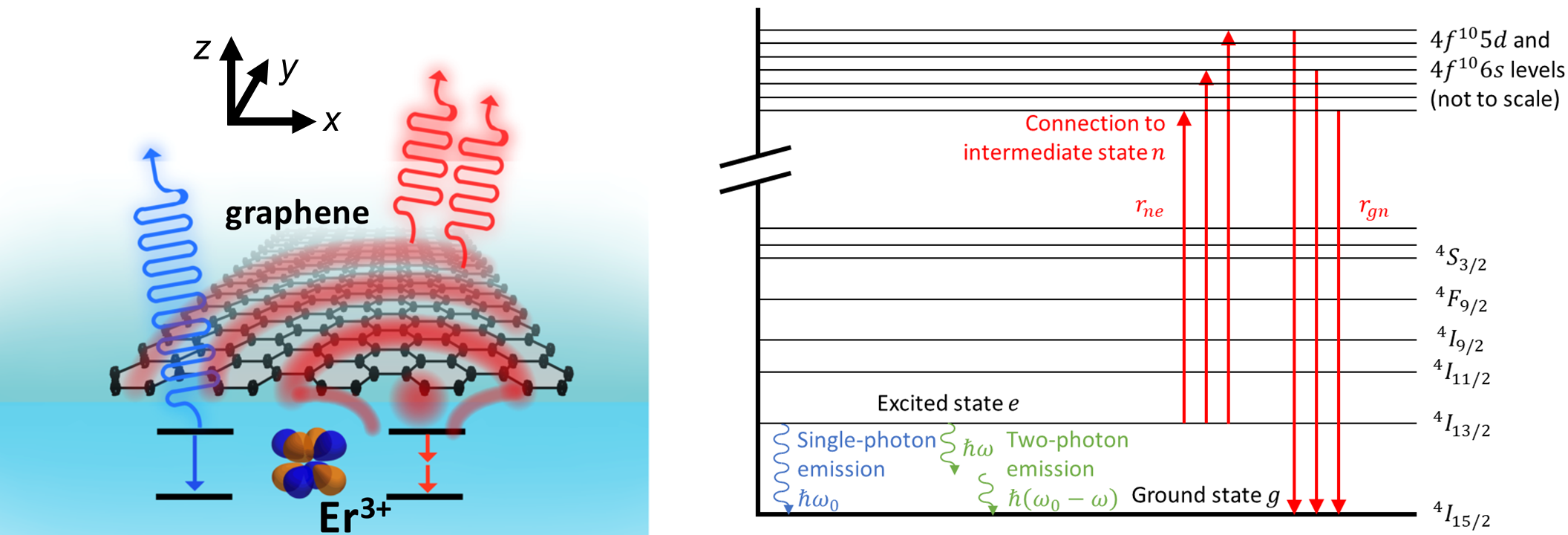}
    \caption{Left: schematic of an emitter beneath a graphene nanoribbon, with single-transition decay shown in blue and 2PSE shown in red. Right: energy levels for the processes involved in two-photon spontaneous emission for Er$^{3+}$. An electron relaxes from the $^4$I$_{13/2}$ state to the $^4$I$_{15/2}$ state, but the matrix elements that determine the process's strength are defined by the electric dipole transitions to the higher 4f$^{10}$5d and 4f$^{10}$6s states.}
    \label{fig:overview}
\end{figure*}

In order to enhance 2PSE to a level that is useful in real-world applications, multiple research efforts have investigated solid state systems --- including quantum wells \cite{Hayat_Ginzburg_Orenstein_2008} and quantum dots \cite{Ota_Iwamoto_Kumagai_Arakawa_2011} --- that permit higher rates of 2PSE. In some experiments, the intermediate state is engineered to exist between the initial and final states, creating a resonant condition in the decay process that enhances the 2PSE transition rate. Other works have considered metallic nanostructures as a method of enhancing 2PSE by supporting plasmonic resonances \cite{poddubny_tailoring_2012}. More recently, it was theoretically predicted that 2D materials that support highly confined optical modes --- including surface plasmons in graphene \cite{Muniz_Manjavacas_Farina_Dalvit_Kort-Kamp_2020, Rivera_Kaminer_Zhen_Joannopoulos_Soljacic_2016} and surface polaritons in hexagonal boron nitride \cite{Rivera_Rosolen_Joannopoulos_Kaminer_Soljacic_2017}  --- could dramatically enhance 2PSE from nearby emitters through the Purcell effect. The underlying premise of those predictions is that the enhancement of single- and two-photon processes scale with the third and sixth powers, respectively, of the confinement factor of an optical mode \cite{Rivera_Kaminer_Zhen_Joannopoulos_Soljacic_2016}. In 2D materials, this confinement factor --- which can be approximated as the ratio between the free-space and surface mode wavelengths --- can be $\sim10^2$, large enough to allow 2PSE rates to approach those of single-transition decay. Indeed, theoretical studies have shown that hydrogen atoms suspended above graphene \cite{Rivera_Kaminer_Zhen_Joannopoulos_Soljacic_2016}, graphene nanodisks \cite{Muniz_Manjavacas_Farina_Dalvit_Kort-Kamp_2020}, and monolayer hexagonal boron nitride \cite{Rivera_Rosolen_Joannopoulos_Kaminer_Soljacic_2017} can exhibit 2PSE rates that become comparable to or even exceed the corresponding single-transition decay processes. However, despite numerous predictions, there has been no experimental evidence of 2PSE occurring in graphene-coupled emitters, even though many works have studied such systems in detail \cite{tielrooij_electrical_2015,cano_fast_2020,lin_distance_2013,federspiel_distance_2015,chen_energy_2010,yeltik_evidence_2013,treossi_high-contrast_2009,guzelturk_near-field_2016,tisler_single_2013,stohr_super-resolution_2012,yu_temperature-dependent_2015,gaudreau_universal_2013,kim_visualizing_2010}. Factors that make observations of 2PSE challenging include: (1) the strong quenching effect of graphene on nearby emitters, driven by non-radiative energy transfer as well as coupling to non-radiative surface modes; (2) the high inefficiencies and low speeds of detectors in the mid- to far-IR, where graphene-mediated 2PSE is likely to occur; (3) the competing processes that drive emitter decay, which can also depend on doping and distance from graphene. Even with these barriers, however, a strongly enhanced 2PSE rate should perturb the overall observed lifetime of a graphene-coupled emitter, and such a perturbation has not been revealed in detailed lifetime measurements.

In this work we perform a comprehensive theoretical analysis of 2PSE from fluorescent rare earth atoms that are optically coupled to graphene and graphene nanoribbons. Our focus on rare earths reflects the fact that they can exhibit high efficiency emission at sufficiently long wavelengths to couple with graphene plasmonic modes which, for achievable carrier densities, are heavily damped at wavelengths $<$1 $\mu$m. Infrared emissions from these ions occur between their 4f electron states, which are located close to the nucleus, causing the emission spectra to remain largely consistent across different host crystals \cite{liu_spectroscopic_2005}. For Er$^{3+}$, photonic crystal designs have been proven effective in enhancing emission sufficiently to observe fluorescence of individual ions \cite{dibos_atomic_2018}. Rare earth doped substrates can also be grown via ion implantation or molecular beam epitaxy, where the placement of dopants can be controlled with nm-precision, and multiple experiments have shown that graphene can affect the emission properties of nearby rare earth ions \cite{tielrooij_electrical_2015,cano_fast_2020}. Here we theoretically investigate both single-transition decay and 2PSE from rare earths near graphene as a number of parameters are varied, including emitter orientation, rare earth species, graphene nanostructure geometry, and graphene doping levels.

A basic illustration of our model system is shown in Figure \ref{fig:overview}, where a rare earth atom --- in this case trivalent erbium (Er\^{3+}) --- is placed near a graphene sheet or nanoribbon. Upon excitation, the atom can decay non-radiatively or radiatively via single-transition decay or 2PSE, emitting photons and/or graphene surface plasmons in the process. All of these processes are affected by the graphene, which provides pathways for non-radiative and radiative decay that are dependent on the graphene carrier density, and also on the graphene geometry. 

In order to precisely quantify the graphene-rare earth interaction, we first determine the relative rates of single-transition decay and 2PSE from the rare earth atoms by numerically calculating the transition probabilities between all intermediate states, which is detailed for Er\^{3+} in Section \ref{sec:sec1basic}. In Section \ref{sec:sec2enhance} we use full field finite difference time domain (FDTD) simulations to calculate the enhancement of both single-transition decay and 2PSE from Er\^{3+} due to both radiative and non-radiative processes enabled by the presence of graphene. Next, in Section \ref{sec:sec3smtm} we explore how graphene affects the 2PSE in divalent samarium and thulium, which have energy levels that are more favorable to 2PSE in comparison to Er\^{3+}. Finally, we discuss how our calculations compare to previous works, and experimental configurations where we predict graphene-enhanced 2PSE to be observable.

\section{Determination of 2PSE rates in rare earths}\label{sec:sec1basic}

Contrary to single-photon emission, in which an ion transitions from an excited state $\hbar \omega_e$ to the ground state $\hbar \omega_g$ by emitting a single photon of energy $\hbar \omega_0 = \hbar (\omega_e - \omega_g)$, 2PSE allows the ion to emit two simultaneous photons, one of energy $\hbar \omega$ and the other $\hbar (\omega_0 - \omega)$. As a second-order process, 2PSE requires a sum over all intermediate states $\hbar \omega_n$ that are connected to the initial and final states by electric dipole selection rules. The total 2PSE rate $\Gamma_0$ can be written in terms of the spectral 2PSE rate $\gamma_0$, which takes the form \cite{Craig_Thirunamachandran}
\begin{multline}\label{eq:eq1}
    \Gamma_0 = \int_{0}^{\omega_0} \gamma_0(\omega) d\omega \\ = \frac{4c}{3 \pi}\alpha^2 k^5 \int_{0}^{1} y^3 (1-y)^3 \times \left| \sum_{n} r_{gn} r_{ne} \right. \\
    \left. \left(\frac{1}{(\omega_n - \omega_e)/\omega_0 + y} + \frac{1}{(\omega_n - \omega_e)/\omega_0 + 1 - y}\right) \right|^2 dy
\end{multline} 
where $r_{gn}$ and $r_{ne}$ represent the electric dipole matrix elements connecting the intermediate energy level to the ground state and excited state, respectively. The variable of integration $y$ represents the fraction of the total emission energy accounted for by the first photon, with the second accounting for $1-y$. A diagram of the intermediate transition processes involved in 2PSE for Er\^{3+} is shown in Figure \ref{fig:overview}, along with the $^4$I$_{13/2}$ $\rightarrow$ $^4$I$_{15/2}$ transition for single-channel decay.

Although the $^4$I$_{13/2}$ $\rightarrow$ $^4$I$_{15/2}$ transition is the dominant emission process in Er\^{3+} within a material host (e.g. Y\_{2}O\_{3}), it is classically electric dipole forbidden due to both states having the 4f$^{11}$ configuration, which has odd parity. The strength of the transition arises from the mixing of these states with higher-lying even parity states in the 4f$^{10}$5d and 4f$^{10}$6s configurations, which occurs due to the host crystal structure breaking the symmetry of the wavefunction \cite{Judd_1962, Ofelt_1962}. Similarly, for 2PSE, the matrix elements $r_{gn}$ and $r_{ne}$ in Eq. (\ref{eq:eq1}) that connect the $^4$I$_{13/2}$ and $^4$I$_{15/2}$ states to other 4f$^{11}$ states will be nonzero; however, the dominant contributions will originate from the electric dipole allowed even-parity (4f$^{10}$) states. To determine $r_{gn}$ and $r_{ne}$, we use Cowan's atomic structure codes to calculate the wavefunctions, diagonalize the matrix, and calculate the resulting transition matrix elements \cite{Cowan_1981}. This allows us to numerically calculate the energy levels for all 4f$^{10}$5d and 4f$^{10}$6s states along with the oscillator strengths for all electric dipole transitions between them and the $^4$I$_{13/2}$ and $^4$I$_{15/2}$ states. The energy levels for the 4f$^{10}$5d and 4f$^{10}$6s levels used for $\omega_n$ in Eq. (\ref{eq:eq1}) are adjusted from the outputs of Cowan's code by adding a constant amount to each, setting the lowest-energy dipole-allowed state to the published value of 73458 cm$^{-1}$ above the ground state \cite{Meftah_Mammar_Wyart_Tchang-Brillet_Champion_Blaess_Deghiche_Lamrous_2016}. Using these matrix elements in Eq. (\ref{eq:eq1}) and a single-photon emission frequency of $\omega_0 = 6500$ cm$^{-1}$ \cite{sardar_absorption_2007}, we compute a spontaneous two-photon emission rate for Er$^{3+}$ of
$ \Gamma_0 = 2.943\times10^{-7}\,\mathrm{s}^{-1}. $
Comparing this value to the experimentally determined single-photon spontaneous emission rate of 125 s$^{-1}$ for Er\^{3+} in Y\_{2}O\_{3} \cite{Weber_1968}, we show that the two processes differ by more than eight orders of magnitude; this relative difference is comparable to what has been predicted for the 2s $\rightarrow$ 1s two-photon transition in hydrogen compared to the 2p $\rightarrow$ 1s single-photon transition \cite{Breit_Teller}.

\section{Enhancement of 2PSE and single-transition decay in Er$^{3+}$ by graphene and graphene nanoribbons}\label{sec:sec2enhance}

While intrinsic 2PSE rates are substantially less than those of single-transition decay, it has been predicted that they can be made comparable by using Purcell enhancement, wherein the decay rate of an emitter is enhanced in the presence of confined optical modes \cite{Rivera_Kaminer_Zhen_Joannopoulos_Soljacic_2016}. This process can occur in any optical cavity, but it can be especially strong in graphene and other 2D materials, which support optical surface waves with wavelengths much shorter than free space. These modes --- which include surface plasmons and phonon polaritons --- enhance single-photon emission rates by a factor that scales with $\eta_0^3=(\lambda_0/\lambda_p)^3$, where $\lambda_0$ is the free-space wavelength and $\lambda_p$ is the wavelength of the surface mode. For two-photon processes, the enhancement scales as $\eta_0^6$. While metals can support surface modes with $\eta_0 \sim 10$, graphene plasmons exhibit $\eta_0 > 100$ \cite{Jablan_Buljan_Soljacic_2009}, leading to substantially more 2PSE enhancement over single-transition decay processes. Moreover, when the graphene is patterned into a nanostructure that supports resonant plasmonic modes, the emission enhancement is further increased by the Q factor of the resonance, it can be made to selectively occur at specific frequencies, and the emitted plasmonic modes can more efficiently out-couple to free space.

In general, the modification of the differential two-photon spontaneous emission rate in the presence of a surface is given by \cite{Muniz_da_Rosa_Farina_Szilard_Kort-Kamp_2019} 
\begin{equation}
    \frac{\gamma(\omega, \textbf{r})}{\gamma_0(\omega)} = \frac{\sum_{i, j} |D_{ij}(\omega, \omega_0-\omega)|^2 P_i(\omega, \textbf{r}) P_j(\omega_0-\omega,\textbf{r})}{\sum_{i', j'} |D_{i'j'}(\omega, \omega_0-\omega)|^2}.
\end{equation}
Here $P_i(\omega, \textbf{r})$ depends on the confinement factor $\eta_0$, the wavenumber $k$, and the surface-emitter distance $z_0$ according to $P_i(\omega, \textbf{r}) \sim \eta_0^3 \text{exp}(-2\eta_0 k z_0)$ \cite{Rivera_Kaminer_Zhen_Joannopoulos_Soljacic_2016} and represents the Purcell enhancement of an electric dipole at location $\textbf{r}$ emitting at frequency $\omega$ and oriented along the axis $i=[x, y, z]$. The matrix $D_{ij}$ depends on the electric dipole matrix elements $\textbf{d}_{ne}$ and $\textbf{d}_{gn}$ connecting the intermediate states $n$ to the excited and ground states, respectively, and is given by
\begin{equation}
    D(\omega_1, \omega_2) = \sum_{n} \left(\frac{\textbf{d}_{ne} \textbf{d}_{gn}}{\omega_n-\omega_e+\omega_1} + \frac{\textbf{d}_{gn} \textbf{d}_{ne}}{\omega_n-\omega_e+\omega_2}\right).
\end{equation}
    
For a hydrogenic ion whose excited and ground configurations are both s orbitals, $D_{ij}$ will be diagonal, as the dipole-allowed intermediate p states are aligned along a specific x-, y-, or z-axis and can only be reached when both transitions have the same polarization. For more complicated ions such as Er$^{3+}$, however, the off-diagonal terms will in general not be zero, and their values may be computed by using the Wigner-Eckart theorem to calculate the irreducible tensor components for each transition \cite{Cowan_1981,eckart1930application,wigner1927einige}. After minor simplifications, we find that the off-diagonal components for Er$^{3+}$ contribute 3/4 as strongly as the diagonal ones, so the differential 2PSE enhancement is given by
\begin{equation}\label{eq:approx_enh}
    \frac{\gamma(\omega, \textbf{r})}{\gamma_0(\omega)} \approx \sum_{i, j} A_{i, j} P_i(\omega, \textbf{r}) P_j(\omega_0-\omega,\textbf{r})
\end{equation}
with
\begin{equation}
    A = \begin{pmatrix} 2/15 & 1/10 & 1/10\\ 1/10 & 2/15 & 1/10\\ 1/10 & 1/10 & 2/15 \end{pmatrix}.
\end{equation}
Further details of this calculation can be found in the Supplemental Material for this paper.

\begin{figure*}
    \centering
    \includegraphics[scale=0.38]{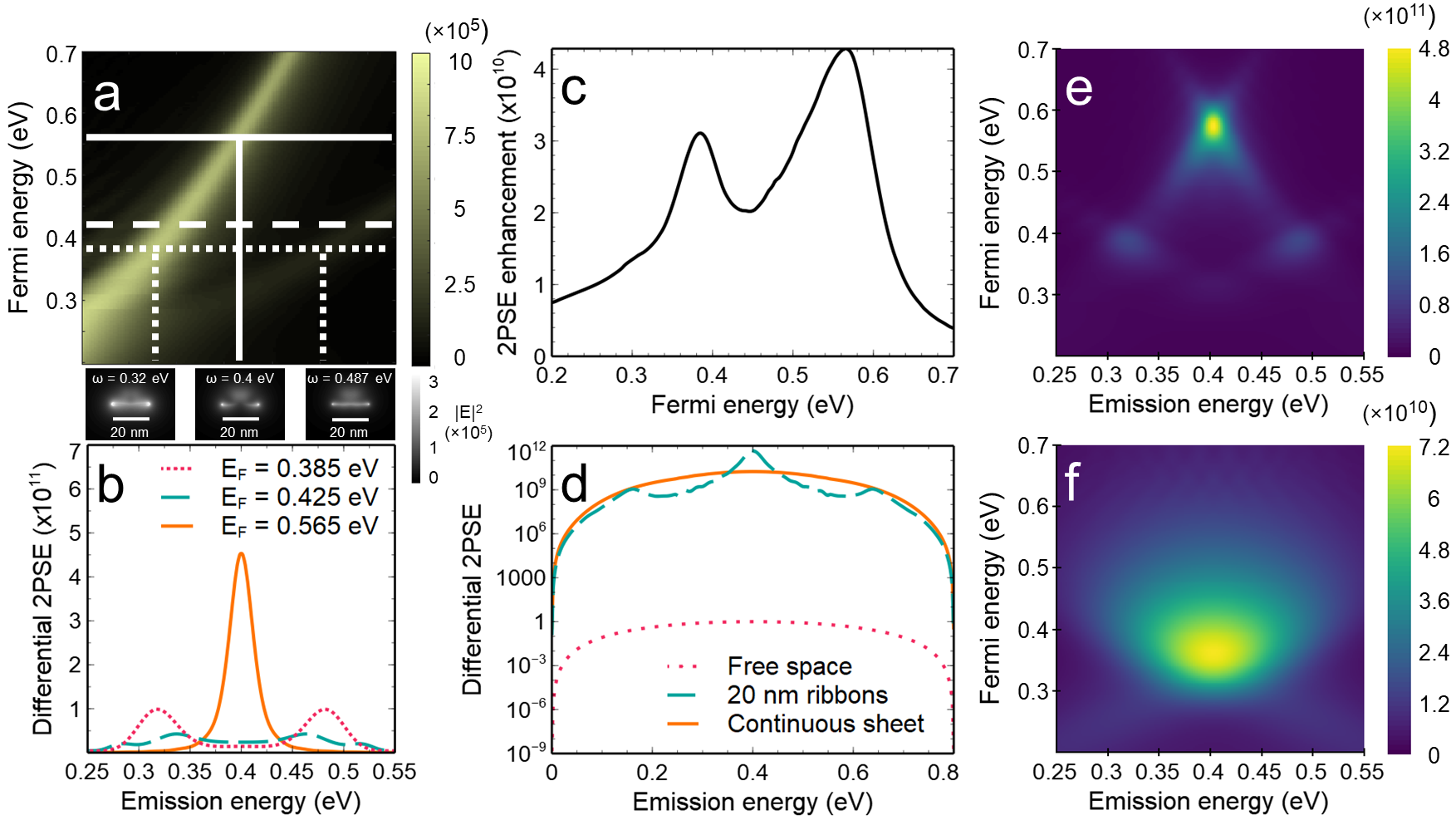}
    \caption{Results for a single dipole orientation. All results consider an Er$^{3+}$ emitter 5 nm away from graphene structures, with dipole moment oriented along the short dimension of the ribbons (or parallel to the sheet). (a) Purcell enhancement for the emitter near 20 nm graphene ribbons. Horizontal lines correspond to the Fermi energies plotted in (b), while vertical lines show the doubly-resonant frequencies for photon emission. Electric field profiles at $E_F=0.385$ eV are shown below, calculated at a lateral distance of 7 nm along the ribbon with the field strength normalized to the value with no graphene present. (b) Differential 2PSE enhancement $\gamma/\gamma_{0, max}$ for the emitter near graphene ribbons for doubly-resonant Fermi energies (pink, dotted; orange, solid) and a non-resonant Fermi energy (blue, dashed). (c) Integrated 2PSE enhancement $\Gamma/\Gamma_0$ for the emitter near 20 nm wide graphene ribbons. (d) Differential 2PSE enhancement $\gamma/\gamma_{0, max}$ at $E_F=0.565$ eV for the emitter with no graphene present (pink, dotted), modified by graphene ribbons (blue, dashed), and modified by a graphene sheet (orange, solid). (e, f) Differential 2PSE enhancement $\gamma/\gamma_{0, max}$ for the emitter near graphene ribbons (e) or a graphene sheet (f). The results in (b), (d), (e), and (f) are normalized by the maximum value of free-space differential 2PSE.}
    \label{fig:differential}
\end{figure*}

Computational methods such as finite difference and finite element methods provide a tool that can be used to analyze the modification of emission rates in a variety of plasmonic environments \cite{akselrod_probing_2014}. To calculate the $P_i$ factors in Eq. (\ref{eq:approx_enh}), we use FDTD (Lumerical) simulations to monitor the power emitted from an electric dipole source placed 5 nm below the graphene as a function of frequency and graphene Fermi level ($E_F$).  We consider the effects of both a continuous graphene sheet as well as ribbons designed in an array with a period equal to three times the ribbon width, with the dipole source positioned beneath the center of the ribbon.  In all simulations, the graphene dielectric properties are modeled using a surface conductivity formalism \cite{Hanson_2008}, with a mobility set to 500 cm$^2$/V$\cdot$s, a value consistent with graphene grown via chemical vapor deposition (CVD) and fitted experimentally in previous measurements of graphene plasmons \cite{hong2022roll,jang2014tunable,kim2018electronically,kobayashi2013production,pirkle2011effect}.

We note that interactions between an emitter and nearby graphene occur through three main mechanisms depending on the Fermi energy to which the graphene is tuned \cite{cox_dipole-dipole_2012,karanikolas_dynamical_2015,tielrooij_electrical_2015,cano_fast_2020,Koppens_Chang_Garcia_de_Abajo_2011,biehs_large_2013,manjavacas_temporal_2012}. When the Fermi energy $E_F$ is less than half the emitter energy, that energy is transferred to graphene by exciting an electron-hole pair. At higher values of $E_F$, the emitter instead transfers most of its energy to free space in the form of propagating photons, or it is damped by Ohmic loss in the graphene sheet. At still higher values of Fermi energy, when $E_F$ exceeds $\sim$70$\%$ of the emission energy, the emission couples to plasmonic modes supported by the graphene. All three can, in principle, contribute to the enhancement of the decay rate; however, the enhancement by plasmonic modes is orders of magnitude larger than the other interaction mechanisms, and when two photons/plasmons are considered, this difference is increased further, such that decay pathways via two-plasmon transitions dominate. We therefore identify the emission enhancement due to graphene as Purcell enhancement throughout this text, although other non-radiative pathways make minor contributions.

In Figure \ref{fig:differential}(a), we plot $P_x(\omega, \textbf{r})$ for the case of a dipole emitter placed 5 nm beneath a 20 nm wide graphene nanoribbon, with the dipole axis oriented along the short axis of the ribbon. The effects of changing the emitter's position beneath the ribbon, the emitter's orientation, or the ribbon's width are all discussed in the Supplemental Material for this paper. For a given value of $E_F$, it is observed that $P_x(\omega, \textbf{r})$ shows two maxima of magnitude $10^5$ - $10^6$, which correspond to the 1\^{st} and 2\^{nd} order plasmonic resonances supported by the graphene nanoribbon, consistent with previous works \cite{Koppens_Chang_Garcia_de_Abajo_2011}.

For 2PSE, emission of one photon at frequency $\omega$ is accompanied by another at $\omega_0 - \omega$, and the total enhancement is determined by the factor $P_i(\omega, \textbf{r}) \times P_j(\omega_0-\omega,\textbf{r})$. When both photons are matched to plasmonic resonances, the 2PSE shows strong enhancement at particular energies; conversely, when one or more photons are not matched to a plasmonic resonance, there is suppression of 2PSE. This behavior is illustrated in Figure \ref{fig:differential}(b), where it is shown that for $E_F = 0.565$ eV there is a strong enhancement of the differential 2PSE near 0.4 eV (3.1 $\mu$m) due to the plasmonic resonance supported in the graphene nanoribbon at that energy, which simultaneously enhances the emission of both 2PSE photons. Likewise, when $E_F$ is tuned to 0.385 eV, plasmonic resonances are supported in the nanoribbon at both 0.32 and 0.49 eV. These support simultaneous emission of photons at both energies which, combined, equal the overall transition energy. This process leads to correspondingly large differential 2PSE enhancements at those photon energies. For an $E_F$ of 0.385 eV, however, the nanoribbons support resonances at 0.34 and 0.53 eV, and thus each resonant emission must be accompanied by a non-resonant process with minimal enhancement, causing the differential 2PSE to be low at both energies. This behavior is further illustrated in Figure \ref{fig:differential}(c), where the wavelength-integrated 2PSE enhancement is plotted as a function of $E_F$, showing maxima when the nanoribbons support plasmonic resonances at energies that can be summed or doubled to equal the overall Er$^{3+}$ transition energy.

The localized resonances in graphene nanoribbons provide energy selectivity to the 2PSE process and also aid in free-space emission; however, their relatively low quality factors ($Q\sim 10$) make it such that they do not enhance 2PSE substantially above what is realized in unpatterned graphene, which supports highly confined plasmonic modes over a large bandwidth.  This is illustrated in Figure \ref{fig:differential}(d,e,f), where the differential 2PSE enhancement is compared between bare graphene and graphene nanoribbons over a range of carrier densities.  Both exhibit enhancements of 10\^{10}-10\^{11}, and while the differential enhancement at $\frac{1}{2}\omega_0 \sim 0.4$ eV can be greater for tuned nanoribbons than unpatterned graphene, the total 2PSE enhancement is comparable for both geometries.  

\begin{figure}
    \centering
    \includegraphics[scale=0.53]{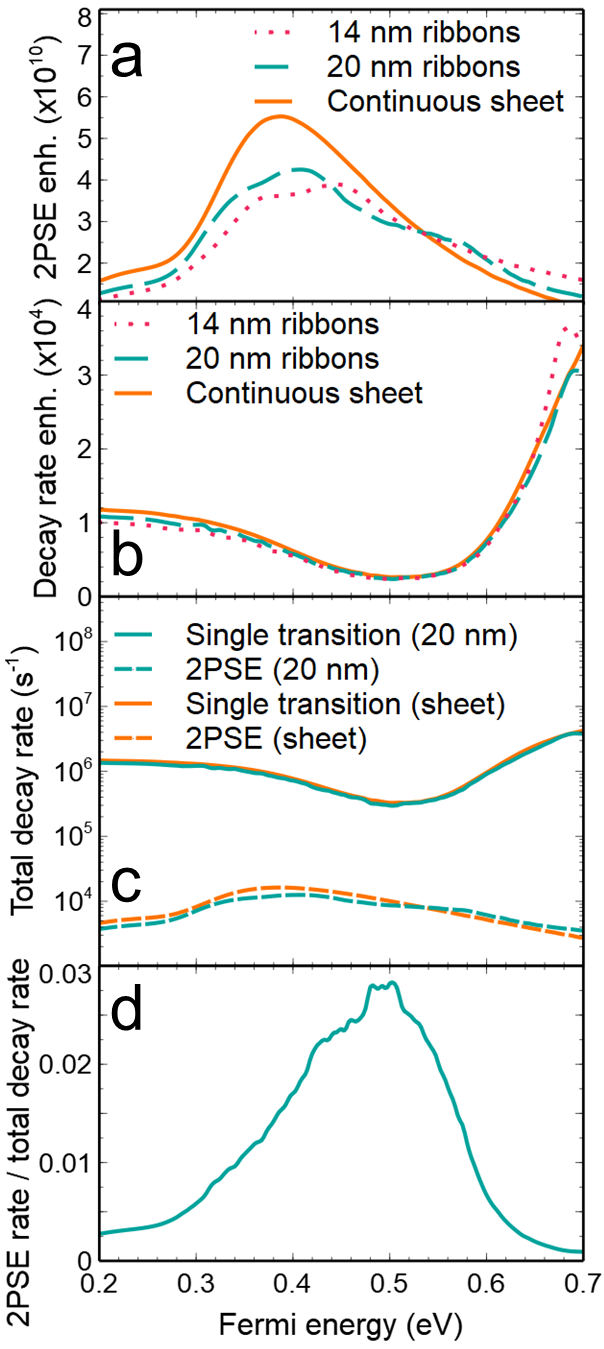}
    \caption{Comparison of single-transition decay and two-photon emission. All results consider an Er$^{3+}$ emitter 5 nm away from graphene structures, with all dipole orientations averaged according to Eq. (\ref{eq:approx_enh}). (a) Enhancement $\Gamma/\Gamma_0$ of total two-photon emission relative to a dipole with no graphene present. (b) Enhancement of single-transition decay at 1538 nm relative to a dipole with no graphene present. (c) Total single-transition decay rate compared to total two-photon emission rate in the presence of 20 nm ribbons and a continuous graphene sheet. (d) Fraction of total decay rate that can be attributed to 2PSE for 20 nm graphene ribbons.}
    \label{fig:overall}
\end{figure}

The above analysis considers only emitters with dipole orientations along the x-axis as defined in Figure \ref{fig:overview}, which most effectively couple to the plasmonic resonances of the ribbon. Aligning the dipole along the y- and z-axes allows for emission into the unbound plasmonic resonances that propagate along the length of the ribbon and behave identically to the unpatterned graphene, and into localized ``dark" mode resonances that have no net dipole moment but can still couple to nearby emitters \cite{Koppens_Chang_Garcia_de_Abajo_2011}. The overall enhancement factors for different dipole orientations do not vary significantly; however, the ``dark" resonances do occur at different frequencies than the resonances observed for an x-polarized dipole (see Supplemental Material). As a result, when all dipole orientations are averaged together (including cross terms), the net effect of the nanoribbons is a Fermi level-dependent 2PSE enhancement that has less well-defined maxima, but that reaches an enhacement of $4\times 10^{10}$ and $5\times 10^{10}$ for the nanoribbons and unpatterned graphene, respectively, as shown in Figure \ref{fig:overall}(a). The lateral position of the emitter also affects the magnitude of the Purcell enhancement (see Supporting Information). Emitters located in the middle of the gap between two ribbons experience approximately an order of magnitude less enhancement than those directly beneath the nanoribbons.  

\begin{figure}
    \centering
    \includegraphics[scale=0.53]{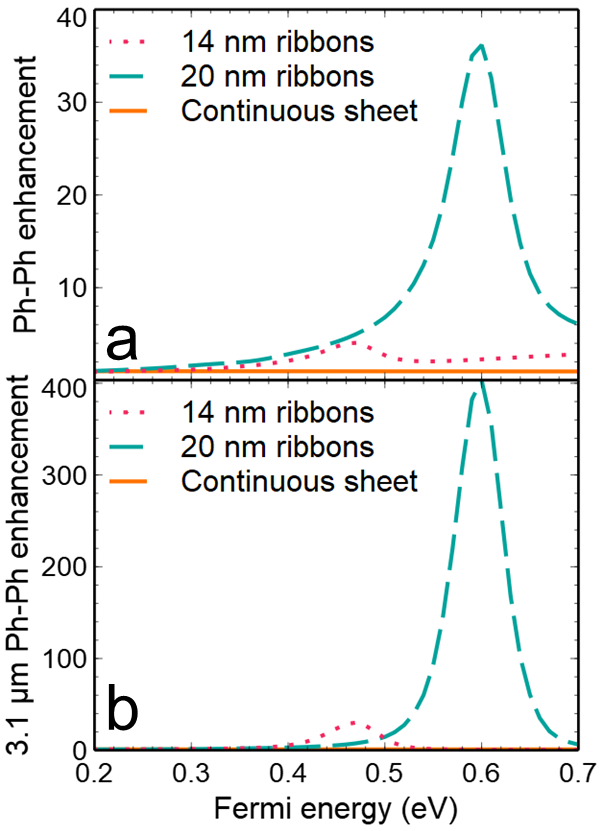}
    \caption{Enhancement of two-photon emission. All results consider an Er$^{3+}$ emitter 5 nm away from graphene structures, with all dipole orientations averaged according to Eq. (\ref{eq:approx_enh}). (a) Enhancement of purely radiative (photon-photon) emission. (b) Enhancement of purely radiative (photon-photon) emission in the 3 - 3.2 $\mu$m range.}
    \label{fig:integrated}
\end{figure}

Although 2PSE is dramatically enhanced in both patterned and unpatterned graphene, the majority of the emission occurs in the form of confined plasmonic modes, which mostly decay non-radiatively.  However, some fraction of the emitted plasmons weakly couple to free space as far-field photons. In Figure \ref{fig:integrated} we calculate the total, Fermi energy-dependent, enhancement of 2PSE from Er$^{3+}$ that results in far-field photon pairs due to continuous graphene and graphene nanoribbons. These results, which consider randomized dipole orientations, were performed by monitoring far-field radiation from a dipole with and without the presence of graphene. The resulting far-field enhancement factors were then utilized in place of the $P_{i}(\omega,r)$ factors in Eq. (\ref{eq:approx_enh}) to obtain the graphene nanoribbon enhancement of two-photon far-field emission. We find only modest enhancements, with the continuous graphene sheet providing no benefit, and the graphene nanoribbons enhancement reaching $\sim 35$.  However, as shown in Figure \ref{fig:integrated}(b), the differential enhancement exceeds 400 when emitting near $\frac{1}{2}\omega_0 \sim 0.4$ eV (3.1 $\mu$m), where the emitted photon pairs have similar energies. Our calculations show that 20 nm wide nanoribbons display substantially more enhancement than 14 nm nanoribbons, despite the latter supporting larger confinement factors in the resonant modes at $\sim 0.4$ eV (3.1 $\mu$m). We attribute this behavior to the better free-space photon-plasmon wavelength matching of the 20 nm nanoribbons. However, we note that even with enhancement factors of 400, the free-space 2PSE is still six orders smaller than conventional single-transition decay for Er$^{3+}$. The degree of free-space emission enhancement can reach $\sim 10^6$ when we instead consider emission in which one decay radiates to the far-field and the other decays non-radiatively, creating a non-entangled, down-converted photon; this behavior is demonstrated in the Supplemental Material of this paper.

While we find the production of free-space photons to be inefficient via graphene-enhanced 2PSE, it is interesting to investigate how it affects the overall lifetime of the Er$^{3+}$.  In order to understand the relative contribution of the 2PSE, it is necessary to also calculate the enhancement by graphene of conventional single-transition decay for the Er$^{3+}$, which is found by calculating the Fermi level-dependent factor $P_i(\omega, \textbf{r})$ at $\omega = 0.8$ eV with the dipole 5 nm from the graphene surface, averaged over all orientations. Figure \ref{fig:overall}(b) shows the results of those calculations for a continuous sheet of graphene, as well as nanoribbbons of 14 nm and 20 nm widths, revealing enhancement factors of $\sim 10^4$, consistent with previously published predictions \cite{Koppens_Chang_Garcia_de_Abajo_2011}. The total emission rates of both single-transition decay and 2PSE can be obtained by multiplying their respective enhancement factors by their free-space rates, which is shown in Figure \ref{fig:overall}(c). These results show that while the 2PSE enhancement is six orders higher than single-transition enhancement for Er$^{3+}$, the overall 2PSE rate is still less than single-transition decay by more than a factor of 10. These results are due to the large difference in initial base rates between single-transition decay and 2PSE, which differ by more than $10^8$. A similar result regarding 14 nm ribbons is shown in the Supplemental Material of this paper. Finally, Figure \ref{fig:overall}(d) shows the Fermi level-dependent fraction of 2PSE compared to the total decay rate of Er$^{3+}$ modified by 20 nm graphene ribbons. It can be seen that the 2PSE can exceed 2.5\% of the total decay at an optimal Fermi energy of $E_F \sim 0.5$ eV.

\section{Enhancement of 2SPE in Sm$^{2+}$ and Tm$^{2+}$}\label{sec:sec3smtm}

The low relative amounts of 2PSE achievable in Er$^{3+}$ via graphene plasmon enhancement are due primarily to the low initial base rates of 2PSE, as well as the large amount of enhancement in competing single-transition decay processes. This motivates the search for other emitters where one or both of those factors are mitigated. To this end, we investigate other rare earth ions that are potentially better suited for realization of a 2PSE-dominant emission source. In particular, one factor contributing to the low base 2PSE rates in Er$^{3+}$ is the large energy spacing between the low-lying 4f$^{11}$ states and the dipole-allowed 4f$^{10}$5d and 4f$^{10}$6s states. This causes the $\omega_n - \omega_e$ term in the denominator of Eq. (\ref{eq:eq1}) to be much larger than the single-photon emission frequency $\omega_0$, reducing the overall 2PSE rate. In divalent rare earth ions, the intermediate states have much lower relative energies above the ground level when compared to their trivalent counterparts \cite{Dorenbos}. We consider two such divalent ions, Sm$^{2+}$ and Tm$^{2+}$. Both of these ions have been synthesized and measured in photoluminescence experiments in fluorides as well as other compounds \cite{Suta_Wickleder_2019}. The first dipole-allowed excited states in these ions are at approximately 24500 cm$^{-1}$ and 25000 cm$^{-1}$ relative to the ground state, respectively, which are lower than the corresponding Er$^{3+}$ level by a factor of more than two. Experimental measurements of Sm$^{2+}$ ions in SrF$_2$ show single-photon emission at 14350 cm$^{-1}$ with a rate of 83.3 s$^{-1}$ when temperature conditions are optimal ($<$45 K) \cite{Alam_Di_Bartolo_1967}. For Tm$^{2+}$ ions in CaF$_2$, the corresponding process occurs at 8966 cm$^{-1}$ with a rate of 278 s$^{-1}$ at liquid nitrogen temperatures \cite{Kiss_1962}. Applying Eq. (\ref{eq:eq1}) to these ions, using oscillator strengths computed through Cowan's code and energy levels adjusted to match the experimentally measured first dipole-allowed level \cite{Martin_Zalubas_Hagan_1978}, we find the free-space 2PSE rates for Sm$^{2+}$ and Tm$^{2+}$ to be 1.928$\times$10$^{-6}$ s$^{-1}$ and 2.386$\times$10$^{-6}$ s$^{-1}$, respectively. These base rates for Sm$^{2+}$ and Tm$^{2+}$ are closer to the experimentally measured single-transition decay rates by factors of 9.8 and 3.6, respectively, in comparison to Er$^{3+}$.

\begin{figure}
    \centering
    \includegraphics[scale=0.53]{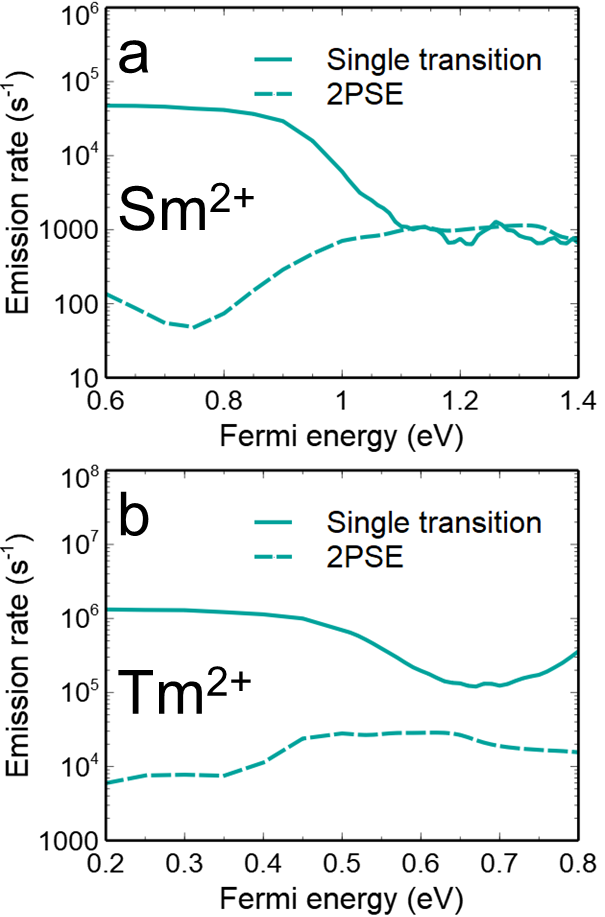}
    \caption{Comparison of total single-transition decay and 2PSE rates for (a) Sm$^{2+}$ and (b) Tm$^{2+}$ when modified by 20 nm graphene ribbons. All results consider an emitter 5 nm away from graphene structures, with all dipole orientations averaged according to Eq. (\ref{eq:approx_enh}).}
    \label{fig:others}
\end{figure}

Next, we explore the relative enhancements by graphene plasmons of 2PSE and single-transition decay in Sm$^{2+}$ and Tm$^{2+}$ using the same methodology described in Section \ref{sec:sec2enhance}. Both the single-transition decay and 2PSE rates for the divalent emitters 5 nm beneath 20 nm graphene ribbons are shown in Figure \ref{fig:others}, computed here using the theoretical single-photon emission frequencies of 15223 cm$^{-1}$ (1.89 eV) for Sm$^{2+}$ and 8421 cm$^{-1}$ (1.04 eV) for Tm$^{2+}$ \cite{Cowan_1981}. These results show that for Sm$^{2+}$, the 2PSE rate can equal the single-transition decay rate, while for Tm$^{2+}$ the 2PSE rate is able to reach nearly 20\% of the single-transition rate. Both ions, however, require higher Fermi levels to achieve a maximal 2PSE fraction.  For Tm$^{2+}$, the optimal Fermi energy of 0.65 eV is beyond what is achievable via electrostatic gating, but can be reached with ionic liquid gating schemes \cite{fang2014active,siegel2021using,tielrooij_electrical_2015}; for Sm$^{2+}$, however, Fermi energies exceeding 1.1 eV are required, which is potentially realizable using chemical doping techniques \cite{khrapach2012novel,song2014iron}.

\section{Discussion}\label{sec:discussion}

Our calculations predict 2PSE rates via graphene plasmon enhancement that are significantly lower than previously published results, which predicted that 2PSE could easily exceed single-transition decay rates when enhanced with surface modes in 2D materials \cite{Muniz_Manjavacas_Farina_Dalvit_Kort-Kamp_2020, Rivera_Kaminer_Zhen_Joannopoulos_Soljacic_2016, Rivera_Rosolen_Joannopoulos_Kaminer_Soljacic_2017}. There are two primary reasons for this difference. First, previous works studied emitter systems with large transition moment lengths ($\textbf{d} = \bra{f} \textbf{x} \ket{i}$), such as the 4s $\rightarrow$ 3s transition of hydrogen. Using the 5p state as the intermediate and incorporating all three dipole polarizations, the hydrogen 4s $\rightarrow$ 3s transition has an average transition moment length of $\sim 2.3$ $\angstrom$ between the two transitions that occur. Such systems have intrinsically large 2PSE base rates, since the 2PSE rate approximately scales as $\textbf{d}^{4}$ \cite{Rivera_Rosolen_Joannopoulos_Kaminer_Soljacic_2017}. For the 4s $\rightarrow$ 3s transition, for example, the 2PSE base rate is $\sim 10^{-2}\,\mathrm{s}^{-1}$. However, coupling an isolated hydrogen atom to a graphene surface that is a few nanometers away is experimentally challenging. In contrast, fluorescent solid state ``color centers" such as those we study here --- which have larger atomic numbers --- have substantially smaller dipole lengths, leading to lower overall 2PSE base rates; for Er$^{3+}$, each intermediate has an average transition moment length of around 0.4 $\angstrom$ or lower, and as shown in Section \ref{sec:sec1basic}, this emitter has a 2PSE base rate of $2.943\times10^{-7}\,\mathrm{s}^{-1}$. Thus, while it is experimentally easier to couple graphene to solid state emitters, the 2PSE enhancement must be much larger in order to be comparable to competing single-transition processes. Second, our calculations use a graphene model that considers a carrier mobility of 500 cm$^2$/V$\cdot$s, which is consistent with what is typical for graphene samples grown via chemical vapor deposition and placed on oxide surfaces \cite{hong2022roll,jang2014tunable,kim2018electronically,kobayashi2013production,pirkle2011effect} but is lower than mobilities considered in previous theoretical works, which assumed mobilities of 10000 cm$^2$/V$\cdot$s \cite{Muniz_Manjavacas_Farina_Dalvit_Kort-Kamp_2020}. The effect of the lower mobility is to increase the single-channel decay rate enhancement, and --- for the graphene nanoribbons --- to decrease the 2PSE enhancement.  This leads to a smaller fraction of ions exhibiting 2PSE in comparison to single-channel processes.

Thus, a main conclusion of this work is that when achievable material systems are considered, the prospects of realizing 2PSE as a dominant emission process are reduced. However, our calculations also indicate the conditions where the effects of 2PSE may be observable in experimentally testable systems. For example, rare earth doped substrates prepared via ion implantation or molecular beam epitaxy can be engineered to place ions at a fixed distance from an overlying graphene sheet. Lifetime-dependent emission measurements of such sample geometries have confirmed modification of the overall single-transition rate of Er$^{3+}$ by both interband and intraband processes in graphene \cite{tielrooij_electrical_2015,cano_fast_2020}. Our calculations indicate that similar experiments can be used to search for evidence of 2PSE by probing the Fermi energy-dependent lifetime of Er$^{3+}$ at 1538 nm, which should be affected by 2PSE near $E_F \sim 0.5$ eV, as shown in Figure \ref{fig:overall}(d). The presence of such a 2PSE contribution to overall decay rates would require lifetime measurements with a time resolution of $\sim$ 1 ns. Moreover, it is critical in such measurements that the  Er$^{3+}$ have a highly uniform distance distribution from the graphene. Previous results were performed using experimental configurations with $\sim$ 1 $\mu$s time resolution and with a broad distribution of distances, such that the signal was dominated by ions located further from the graphene, which are more likely to emit radiatively than through a plasmonic surface mode \cite{tielrooij_electrical_2015,cano_fast_2020}. We note that such measurements would also be aided by measuring high mobility samples, where the substrate is treated or engineered to minimize charge disorder. Larger mobilities would suppress the enhancement of single-transition processes and allow 2PSE to play a larger relative role, making detection easier.

Direct detection of photons emitted from rare earth ions via 2PSE is also possible; however, only a small fraction ($\sim 10^{-9}$) of plasmon pairs generated via 2PSE are emitted to free space as photon pairs. In the limit of high excitation rates, the number of emitted photon pairs in the 3 - 3.2 $\mu$m wavelength range can be estimated to be 222 photon pairs per second. This assumes a Y$_2$O$_3$ substrate with a doped plane of 0.2\% Er$^{3+}$ that is 5 nm below an array of 20 nm graphene nanoribbons with $E_F=0.6$ eV and with collection over a 10 $\mu$m $\times$ 10 $\mu$m area. These numbers can be improved by using larger dopant concentrations and collection areas, and larger graphene nanoribbons which have lower Purcell enhancements but more effectively couple plasmonic modes to free space. However, when emitters are considered that are placed at random lateral positions relative to the graphene nanoribbons, the overall 2PSE will be reduced.

Another possible route to verify our theoretical results is by probing two-photon absorption, the inverse process to 2PSE. In a two-photon absorption (2PA) measurement, two photons with frequencies summing to $\omega_0$ are incident on an atomic system simultaneously, elevating an electron to a higher energy level from which it can decay \cite{lin2010enhanced}. The efficiency of this process is also dependent on $P_i(\omega, \textbf{r}) \times P_j(\omega_0-\omega,\textbf{r})$, which, for single-frequency excitation, must satisfy the condition $\omega = \omega_{0}/2$. Thus, the large Purcell enhancements supported by graphene will also drive large 2PA, which can be monitored by measuring fluorescence at $\omega_0$. For graphene systems coupled to rare earth ions, it would be required to perform 2PA measurements on nanoribbons or some other patterned nanostructure, as the plasmonic modes in a continuous sheet would not couple to free-space excitation.

Our calculations also point to the importance of utilizing emitters with large initial 2PSE base rates for experiments whose goal is to boost 2PSE via Purcell enhancement  to a significant fraction of overall emission. While previous works showed that atomic hydrogen can achieve this condition, the experimentally viability of such an approach is low.  Nevertheless, other emitter systems could be considered, including fluorescent dyes such as ruthenium complexes and rhodamine 123, and fluorescent biomolecules, such as green fluorescent protein chromophore. These emitters are expected to have transition dipole moments corresponding to lengths of 1.1 - 1.3 $\angstrom$ \cite{stark_solute-solvent_2019}, 1.7 $\angstrom$ \cite{chung_determining_2016}, and 1.4 $\angstrom$ \cite{bublitz_electronic_1998}, respectively. Alternatively, our results show that a detailed consideration of the intermediate energy levels allows for identification of emitters with larger 2PSE base rates. Specifically, we find that emitters with intermediate energy levels that are near the excited state, such as divalent rare earths, have larger 2PSE rates which, via Purcell enhancement, can approach single-transition rates.

\section{Conclusion}

We have calculated the 2PSE base rates for Er$^{3+}$, Sm$^{2+}$, and Tm$^{2+}$, and, using full field simulations, determined the degree to which Purcell enhancement from nearby graphene and graphene nanoribbons can enhance overall 2PSE. We show that the overall 2PSE rate enhancement of Er$^{3+}$ ions can exceed $4 \times 10^{10}$ and that it is strongly dependent on the graphene Fermi level. Moreover, we show that the production of entangled free-space photon pairs near 3.1 $\mu$m via 2PSE can be enhanced by 400 by graphene nanoribbons. In contrast to previous studies of hydrogen, we find that the 2SPE rate for Er$^{3+}$ does not become a significant fraction of the overall decay, and we attribute this discrepancy to the low initial 2PSE base rate for Er$^{3+}$ and the lower graphene mobility used in our calculations. These two factors, however, represent realistic experimental conditions, and we outline ways in which 2PSE could be observed in Er$^{3+}$-doped substrates coated with graphene; such systems have already been explored experimentally in regimes where single-transition processes are affected by graphene. Finally, we find that controlling the emitter species can affect the 2PSE contribution to overall emission. Specifically, we show that divalent rare earths that have intermediate states close in energy to the initial excited state have higher overall base rates of 2PSE, and we find that for Sm$^{2+}$ and Tm$^{2+}$, Purcell enhancement can be used to make 2PSE a large overall fraction of the decay.

These findings provide specific guidance for future experiments that aim to generate photon pairs via 2PSE using Purcell enhancement. Moreover, they can be used to engineer next-generation devices that produced entangled photons at high rates for quantum information science applications.

\section{Acknowledgments}

This material is based upon work supported by the U.S. Department of Energy Office of Science National Quantum Information Science Research Centers as part of the Q-NEXT center, which supported the work performed by C. Whisler and V. W. Brar. G. Holdman was supported by a Defense
Advanced Research Projects Agency Young Faculty Award
(YFA D18AP00043). D. D. Yavuz was supported by the Vilas Associates Award of the UW-Madison.

\bibliography{main}

\end{document}